\documentstyle[epsfig,a4,12pt]{article}
\epsfverbosetrue

\def\nuebar{\rm{\bar{\nu_e}}}
\def\nue{\rm{\nu_e}}
\def\s2tw{\rm{ sin ^2 \theta _W }}
\def\munu{\rm{\mu_{\nu}}}
\def\mub{\rm{\mu_B}}
\def\enu{\rm{E_{\nu}}}
\def\emax{\rm{E^{\nu}_{max}}}
\def\hl{\rm{\tau_{\frac{1}{2}}}}
\def\rnusp{\rm{\phi ( \bar{\nu_e} ) }}
\def\th232{\rm{ ^{232} Th }}
\def\cs137{\rm{^{137} Cs }}
\def\ba133{\rm{^{133} Ba }}
\def\u238{\rm{ ^{238} U }}
\def\ur235{\rm{^{235} U}}
\def\241pu{\rm{^{241} Pu}}
\def\pu239{\rm{^{239} Pu}}
\def\dtotal{\rm{\delta_{total}}}
\def\dmeas{\rm{\delta_{det}}}
\def\dsm{\rm{\delta_{SM}}}
\def\dmm{\rm{\delta_{MM}}}

\begin{document}
\hspace*{1cm} \hfill \today

\begin{center}
\LARGE
\bf{
Sensitivities of Low Energy\\
Reactor Neutrino Experiments\\
}

\vspace*{0.3cm}

\Large

Hau-Bin Li~$^{a,b}$
and
Henry T. Wong~$^{a,}$\footnote{Corresponding~author:
Email:~htwong@phys.sinica.edu.tw;
Tel:+886-2-2789-9682;
FAX:+886-2-2788-9828.}
\end{center}

\large

\begin{flushleft}
$^a$ Institute of Physics, Academia Sinica, Taipei 11529, Taiwan.\\
$^b$ Department of Physics, National Taiwan University, 
Taipei 10617, Taiwan.\\
\end{flushleft}

\vspace*{0.3cm}

\normalsize

\begin{center}
{\bf
Abstract
}
\end{center}

The low energy part of the reactor neutrino
spectra has not been measured experimentally.
Its uncertainties 
limit the sensitivities in certain reactor neutrino
experiments.
This article discusses
the origin of these uncertainties and examines
their effects on the measurements of
neutrino interactions with electrons and nuclei. 
The discrepancies between
previous results and 
the Standard Model expectations
can be explained by the under-estimation of the
reactor neutrino spectra at low energies.
To optimize the experimental sensitivities,
measurements for $\nuebar$-e cross-sections
should focus on events with
large ($>$1.5~MeV) recoil energy while those
for neutrino magnetic moment searches 
should emphasize on events $<$100~keV.
The merits and attainable accuracies 
for neutrino-electron scattering experiments 
using artificial neutrino sources are discussed.

\vspace*{0.2cm}

\begin{flushleft}
\end{flushleft}

\clearpage

\normalsize

\section{Introduction}

Nuclear power reactors are intense $-$ and 
readily available $-$  sources
of electron anti-neutrinos ($\nuebar$)
at the MeV energy range. 
They remain important tools in the
experimental studies of neutrino properties and
interactions. 

Reactor-based
neutrino oscillation
experiments~\cite{nuphys,nu2000} 
typically rely on
the interactions between
$\nuebar$ and proton, usually in the form
of hydrogen in liquid scintillator:
\begin{equation}
\label{eq::nup}
\rm{ 
\nuebar ~ + ~  p ~ \rightarrow ~ e^+ ~ + ~ n ~~ .
}
\end{equation}
With an interaction threshold of 1.8~MeV and
a typical positron detection threshold of $>$ 1~MeV,
the ``reactor neutrino spectrum'' [$\rnusp$]
above 3~MeV has to
be known to derive the physics results. 

There are by-now standard procedures for evaluating
$\rnusp$ using the
reactor operation
data. An accuracy of up to 1.4\% 
between calculations and measurement
has been achieved in the 
integrated flux~\cite{bintflux}. 
The measured
differential spectrum from the Bugey-3 experiment~\cite{bugey3}
was compared with three models of deriving
$\rnusp$.  The best one
gave an accuracy of better than 5\% from 2.8~MeV to
8.6~MeV neutrino energy, while the other two gave
discrepancies at the 10-20\% level in part of this
energy range.

The conclusion of these studies is that $\rnusp$
above 3~MeV can be considered to
be calculable to the few \% level. 
Therefore, long-baseline
reactor neutrino oscillation experiments 
do not require  ``Near Detectors'' for the flux normalization
purpose.
These experiments,
including
Chooz and Palo Verde which have been
performed, KamLAND in operation and Borexino under construction,
focus on the large mixing angles (big oscillation
amplitudes) and small $\rm{\Delta m^2}$
parameter space. 

However, 
the $\rnusp$ below 3~MeV 
had neither been measured experimentally, nor
thoroughly addressed theoretically. 
In Section~\ref{sect::rnusp},
we summarize the essence of
the calculations of reactor neutrino spectra,
and describe the origins of the uncertainties
at low energy.
The potential contributions of these effects to
the accuracies of experimental measurements are discussed.
In particular, we investigate the case of
neutrino-electron ($\nuebar$-e) scatterings
in Section~\ref{sect::nue}, and show how
the uncertainties in the low energy part
of the reactor neutrino spectra 
limit the sensitivities in
the cross-section measurements as
well as the search of neutrino magnetic moments.
The effects on the study of neutrino interactions
on nuclei are discussed in Section~\ref{sect::nunuclei}.
A list of relevant cross-sections scattered in the
literature are compiled.

\section{Reactor Neutrino Spectra}
\label{sect::rnusp}

Electron anti-neutrinos are emitted in
a nuclear reactor through $\beta$-decays
of unstable nuclei produced by the fission
of the four major fissile elements in
the fuel: $\ur235$, $\u238$, $\pu239$, $\241pu$.
Hundreds of different daughter nuclei are involved,
each having its own decay life-time,
branching ratio and
Kurie distribution,
some of which are not completely known.
The inputs for calculating the
overall $\rnusp$ 
are derived from
two alternative approaches:
(I) modelings on the level densities
and nuclear effects~\cite{thrnu1,avignone}, or
(II) measurements of $\beta$-spectra
due to neutron hitting the fissile
isotopes~\cite{bill}.
The Bugey-3 experiment compared their
data with these approaches~\cite{bugey3} and concluded
that the predictions of $\rnusp$ from (II)
are consistent
with measurements at the $<$5\% level
for $\enu \sim$2.8-8.6~MeV.
The agreement between Approach (I) and data as well as among the two approaches
are typically  at the 10\% level, and can deviate to the 20\%
level at part of the energy range.
The Bugey-3 results contradicted those of Ref.~\cite{nupruss}
which claimed a discrepancy with (II) by 10\%
in spectral shape.
In addition, recent studies on
the non-equilibrium effects~\cite{nupeq} suggested
that the corrections may be as large as
25\% for reactor neutrino oscillation experiments.

\begin{figure}
\centerline{
\epsfig{file=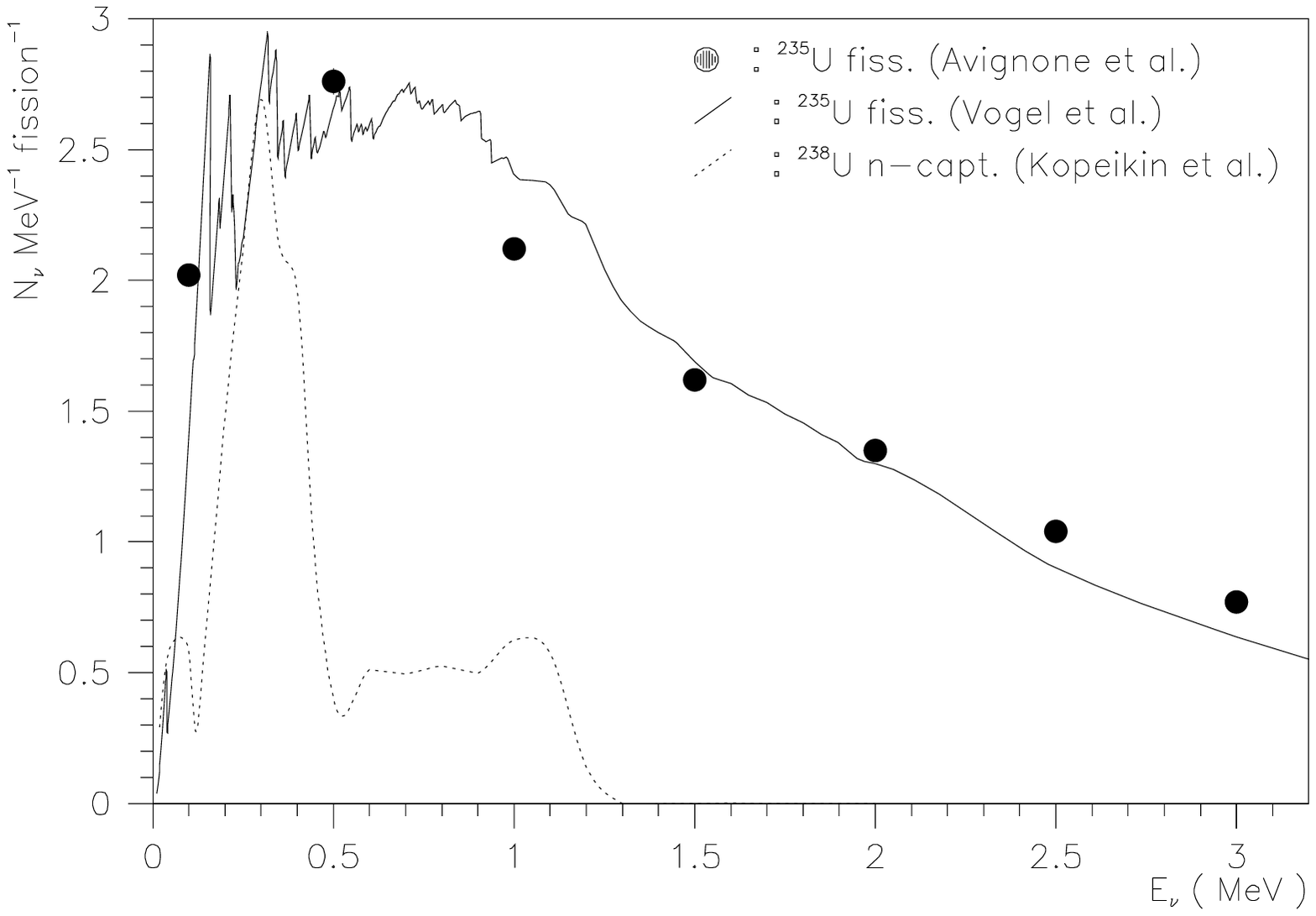,width=14cm}
}
\caption{
Calculated low energy reactor neutrino spectra
due to fission
from Refs.~\cite{avignone} and \cite{vogelengel},
and that due to $\beta$-decays following
neutron capture on $^{238}$U from Ref.~\cite{russcs}.
}
\label{lowrnuspec}
\end{figure}

All these intensive and comprehensive efforts were
motivated by oscillation studies
with proton target~\cite{nupcs,russcs} via the
interaction channel given in Eq.~\ref{eq::nup}.
They may be inadequate for the other
experiments with reactor neutrinos.
In particular, there are no measurements
as well as systematic theoretical efforts on
$\rnusp$ below $\enu \sim$2.8~MeV.
Existing compilations come from 
Refs.~\cite{avignone} and \cite{vogelengel}
which were derived from the summations of $\beta$-decays 
of fission fragments using different modelings.
The calculated spectra below 3~MeV are displayed
in Figure~\ref{lowrnuspec}, showing
that the models are consistent among themselves
only to the 20\% level.
The discontinuities in the spectrum
of Ref.~\cite{vogelengel} 
are due to 
the end-point effects of the many $\beta$-spectra.

There are many additional effects necessary 
to be considered
in order to perform a correct evaluation
of $\rnusp$ at low energies.
There are many more $\beta$-decays with Q-values less
than 3~MeV that have to be modeled on with Approach~(I).
Similarly, the measurements with Approach~(II) were performed
with an exposure time for neutrons on $\ur235$
of 15~hours which is sufficient only to 
bring the  $\beta$-activities above 3~MeV
into equilibrium, and the measured $\beta$-spectra
had a threshold of 2~MeV kinetic energy for the electrons.
Consequently, fission products with live-times longer than 10~hours
and $\beta$-decays with end-points less than 2~MeV
are not accounted for.
Examples of fission daughters belonging to
this category include:
$\rm{^{97}Zr}$ ($\rm{\emax = 1.92~MeV}$; $\rm{\hl = 17~h}$),
$\rm{^{132}Te}$ ($\rm{\emax = 2.14~MeV}$; $\rm{\hl = 78~h}$) and
$\rm{^{93}Y}$ ($\rm{\emax = 2.89~MeV}$; $\rm{\hl = 10~h}$).

The treatment is even more
complicated for those with life-times comparable
to a reactor cycle (12 to 18 months).
During a typical reactor shut down, only
a fraction of the fuel elements are replaced, and
the spent fuel is temporarily stored in the water tank
within the reactor building (that is, in
the vicinity of the experimental site). 
The old and new
fuel elements are usually re-oriented within
the core for the next cycle.
The complications can be illustrated with
a notable example. The fission daughter $^{90}$Sr 
has a half-life of 29.1~y and a cumulative yield
of 5.4\% per $\ur235$-fission~\cite{fptab}. 
It would give
rise to two subsequent $\beta$-decays with maximum $\enu$
of 0.55~MeV and 2.27~MeV, respectively.
Other examples include:
$\rm{^{106}Ru}$ ($\rm{\emax = 3.54~MeV}$; $\rm{\hl = 372~d}$)
and 
$\rm{^{144}Ce}$ ($\rm{\emax = 3.00~MeV}$; $\rm{\hl = 285~d}$).

Neutrons produced in fission can be absorbed
by the fission fuel elements as well as by the
surrounding materials. Some of the final
states are unstable and can give rise to $\beta$-decays
producing $\nuebar$. There is also a smaller
fraction of $\nue$'s due to isotopes which decay
via electron captures and $\beta ^+$-emissions~\cite{nuereactor}.
Almost all such processes have Q-values below
3~MeV and hence contribute only to the low energy
part of $\rnusp$. 
The major contribution from this 
category~\cite{russcs,ncapfis} is
expected to be from the reaction
$\rm{ ^{238} U ( n , \gamma ) ^{239}U }$,
leading to  subsequent $\beta$-decays
via\\
$\rm{
^{239}U ( T_{\frac{1}{2}}=23~min; \emax = 1.26 ~MeV )
\rightarrow  
^{239}Np ( T_{\frac{1}{2}}=2.4~days; \emax = 0.71 ~MeV )
\rightarrow
^{239}Pu ~ .  
}$\\
The typical $\rnusp$ spectrum from the decays of $^{239}$U
is displayed in Figure~\ref{lowrnuspec}, showing
that the flux is comparable
to those due to $\beta$-decays of fission daughters.
The complicated
non-equilibrium effects 
for both Reactor ON and OFF periods~\cite{nonequil} 
from the various neutron capture channels
will contribute further
to the uncertainties in the description of
$\rnusp$ at low energies.
Meticulous book-keeping and complicated
calculations are necessary to account 
for the various effects.

All these processes have not been quantitatively
addressed.
In addition, errors in the evaluation of $\rnusp$
tend to be
under-estimations due to some physical
processes not accounted for.
This would give rise to an excess of events
which may mimic positive signatures for
anomalous effects.
Therefore, one should be cautious about
the conceptual design 
of experiments and the interpretation of data
where the low energy reactor $\nuebar$ 
plays a role,  such as in
neutrino-electron scatterings.
Further work on the calculations 
of $\rnusp$ at low energies
and demonstrations of their accuracies 
would be of interest.

\section{Neutrino-Electron Scatterings}
\label{sect::nue}

Experiments on
neutrino-electron ($\nuebar$-e) scatterings
\begin{equation}
\rm{ 
\nuebar ~ + ~  e^- ~ \rightarrow ~ \nuebar ~ + ~ e^- ~~ .
}
\end{equation}
provide measurements of Standard Model 
parameters ($\rm{g_V}$,$\rm{g_A}$)
for electroweak interactions, as well as a 
probe to study the interference effects 
between the charged and neutral currents~\cite{kaiser}.
The process is also a sensitive way to
study the electromagnetic form factors
of neutrino interactions with the photons, 
and in particular, the  
neutrino magnetic moments~\cite{vogelengel}.
The interaction vertex probed 
is the same as that giving rise to
neutrino radiative decays~\cite{rdk}:
$\rm{
\nu_1  ~ \rightarrow ~ \nu_2 ~ + ~ \gamma ~~ ,
}$
providing sensitivities competitive even to
the limits derived in supernova SN1987a~\cite{sn1987a}.

\begin{figure}
{\bf (a)}
\centerline{
\epsfig{file=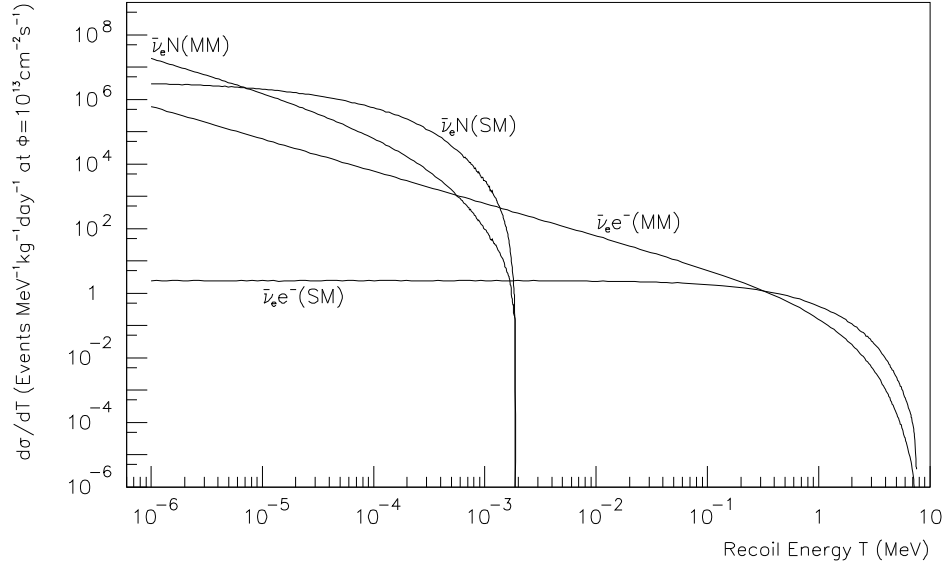,width=13cm}
}
{\bf (b)}
\centerline{
\epsfig{file=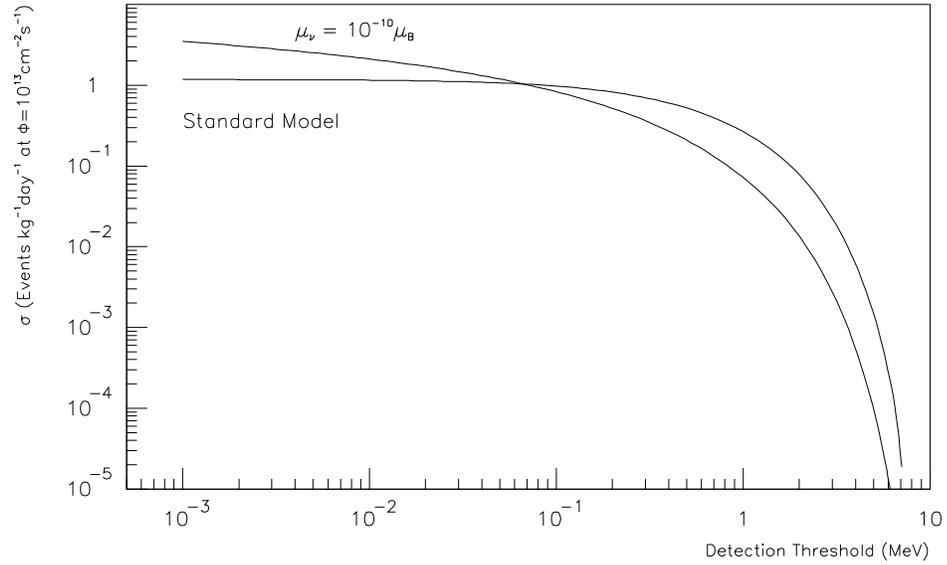,width=13cm}
}
\caption{
(a)
Differential cross section showing the
recoil energy spectrum in
$\nuebar$-e and coherent $\nuebar$-N 
scatterings,
at a reactor neutrino flux of 
$\rm{10^{13}~cm^{-2} s^{-1}}$,
for the Standard Model processes and
due to a neutrino
magnetic moment of 10$^{-10}~\mub$.
(b)
The integral event rates as a function
of the detection threshold of the
recoil electrons in the $\nuebar$-e
processes.
}
\label{nuerecoil}
\end{figure}

The experimental observable is the kinetic energy of the
recoil electrons(T). 
Following Ref.~\cite{vogelengel}, the differential cross section is
given by :
\begin{equation}
\rm{
\frac{ d \sigma }{ dT } ( \nu - e ) ~ =  ~
( \frac{ d \sigma }{ dT } ) _{SM}  ~ + 
( \frac{ d \sigma }{ dT } ) _{MM}  ~ .
}
\end{equation}
The Standard Model(SM) term is 
\begin{equation}
\label{eq::sm}
\rm{
( \frac{ d \sigma }{ dT } ) _{SM}  ~ = ~ 
\frac{ G_F^2 m_e }{ 2 \pi }
[ ( g_V + g_A )^2 + ( g_V - g_A )^2  [ 1 - \frac{T}{E_{\nu}} ]^2
+ ( g_A^2 -  g_V ^2  ) \frac{ m_e T }{E_{\nu}^2}  ] 
}
\end{equation}
where
 $\rm{ g_V = 2 ~ \s2tw - \frac{1}{2}}$ and 
$\rm{ g_A =  - \frac{1}{2}}$
 for $ \nu_{\mu}$-e and $\nu_{\tau}$-e scatterings
which proceed via neutral-current interactions only,
while $\rm{g_V \rightarrow g_V + 1}$ and
 $\rm{g_A \rightarrow g_A + 1}$ for $\nue$-e scatterings,
where both charged and neutral currents are involved.
The expression can be modified for 
$\bar{\nu}$-e scatterings by the replacement
$\rm{ g_A \rightarrow - g_A }$ to account for the
effects due to different helicities.
The magnetic moment(MM) term is 
a non-Standard Model process given by
\begin{equation}
\label{eq::mm}
\rm{
( \frac{ d \sigma }{ dT } ) _{MM}  ~ = ~
\frac{ \pi \alpha _{em} ^2 \munu ^2 }{ m_e^2 }
 [ \frac{ 1 - T/E_{\nu} }{T} ]
}
\end{equation}
where the neutrino 
magnetic moment $\munu$ is often expressed
in units of the Bohr magneton($\mub$).
The process can be due to 
{\it diagonal} and {\it transition}
magnetic moments, 
which change only the spins and 
both the spins and flavors, respectively.
The MM term has a 1/T dependence and hence dominates
at low electron recoil energy. The expected recoil
differential spectra for both processes
are depicted in Figure~\ref{nuerecoil}a.
At energy transfer comparable to the inner-shell
binding energies of the target,
a small and known correction factor 
has to be applied to the cross-section formulae~\cite{nuebind}.
The spectra below 2~keV are due to the 
neutrino coherent scatterings on nuclei~\cite{nuphys,coher},
derived from Eqs.~\ref{eq::cohsm} and \ref{eq::cohmm}
to be discussed further in Section~\ref{sect::nunuclei}. 

Experimentally, the interactions of 
$\nu_{\mu}$-e/$\bar{\nu_{\mu}}$-e~\cite {charm2}
and $\rm{\nu_e}$-e~\cite{lanl}
have been studied with high energy and intermediate
energy accelerator neutrinos.
Although $\nuebar$-e have 
been observed with reactor neutrinos~\cite{reines,kurt,rovno},
the MeV-energy range is still a relatively untested range
where there are still big uncertainties in the measured
cross-sections.
Indeed, the results of Ref.~\cite{reines} give
rise to different levels of consistencies (or slight
discrepancies) with the 
Standard Model expectations when 
different $\rnusp$ were used~\cite{avignone,vogelengel,reines,kyuld}.
There are various 
current experiments~\cite{munu,texono,russ}
pursuing this subject.

\subsection{Reactor Neutrinos: $\nuebar$-e Scatterings}

In this Section, we investigate the effects of
uncertainties in $\rnusp$  to
the sensitivities of SM cross-section measurements and 
limits of MM searches. 
The prescriptions for evaluating
$\rnusp$ from Ref.~\cite{vogelengel} were adopted.
Since only the relative errors are considered,
the conclusions would be independent of the 
fine details of the models used.
The uncertainties in $\rnusp$
were parametrized by two variables
$\xi$ and $\Delta$ such that the
spectra above and below $\xi$ are taken 
to be accurate to 5\%
and $\Delta$\%, respectively.

\begin{figure}
{\bf (a)}
\centerline{
\epsfig{file=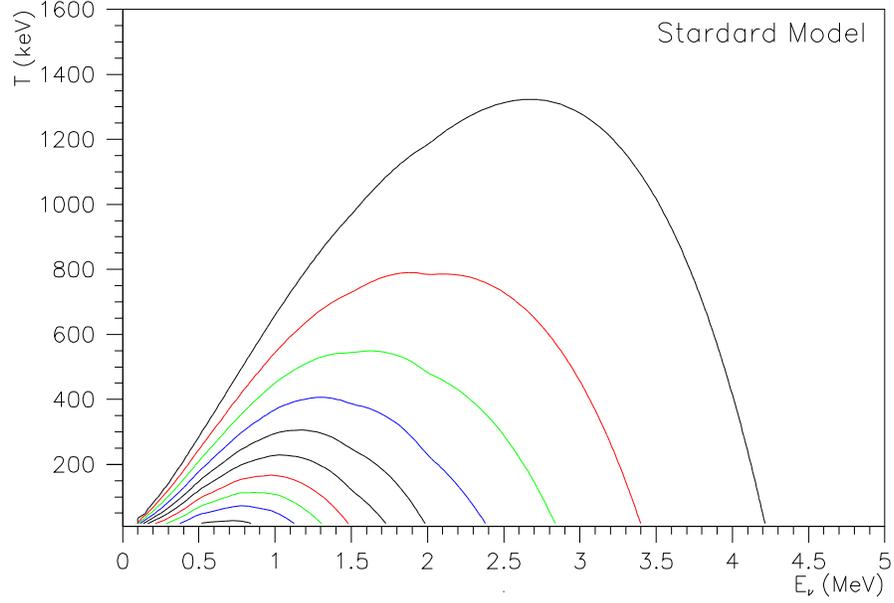,width=12cm}
}
{\bf (b)}
\centerline{
\epsfig{file=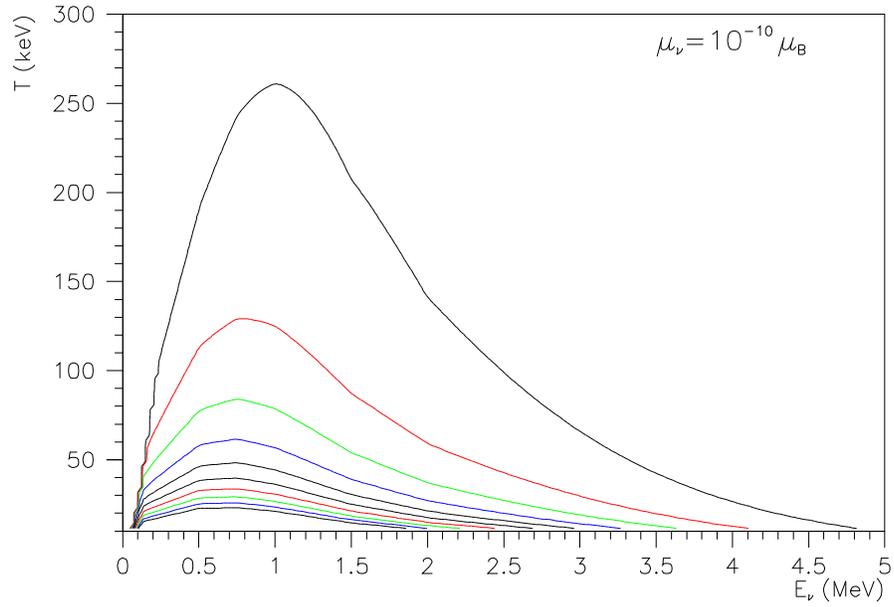,width=12cm}
}
\caption{
Correlation plots for recoil energy (T) versus
neutrino energy ($\enu$)  for (a) Standard Model and  
(b) Magnetic Moment contributions 
due to $\nuebar$-e scattering
from the reactor.
Adjacent contours represent 
equipartition levels of event rates normalized
to the innermost contour.
Magnetic scatterings are due mostly to low energy
neutrinos giving rise to low energy electron recoils.
}
\label{TvsEnu2D}
\end{figure}

The correlations between electron recoil energy (T) and
neutrino energy ($\enu$) for
both SM and MM 
processes (Eqs. \ref{eq::sm} and \ref{eq::mm}, respectively)
due to $\rnusp$ are displayed in
Figure~\ref{TvsEnu2D}a and \ref{TvsEnu2D}b, respectively.
Both distributions peak at small $\enu$ and T.
The contours represent 
equipartition levels of event rates
normalized to the largest value
at the innermost contour.
It can be seen that 
(1) most $\nuebar$-e events for both SM and MM are of 
low recoil energies, and 
(2) they are mostly due to interactions by 
low energy neutrinos,
the contributions of   which are
more pronounced in MM than in SM.
As illustrations, 84\%, 64\% and 29\% of the $\nuebar$-e
SM scattering events at 100~keV recoil
energy are due to neutrinos with energy
less than 3, 2 and 1~MeV, respectively.  

The overall accuracy ($\dtotal$)
in a typical reactor experiment
depends on: 
(a) the contributions from  
the uncertainties
of $\rnusp$ to the SM ($\dsm$) and MM ($\dmm$)
cross-sections, and 
(b) the measurement uncertainties ($\dmeas$)
which include the combined effects
of the experimental systematic and statistical errors,
including those introduced
in Reactor ON$-$OFF subtraction.
In the case where the MM contributions
are negligible, one can write
\begin{equation}
\rm{
\dtotal ^2 ~ = ~
\dmeas ^2 ~ + ~ \dsm ^2 ~ .
}
\end{equation}

To achieve reasonable statistical accuracy,
most experiments compare data 
above a certain detection threshold
with the integrated cross-sections.
The integral recoil spectra for different
threshold is shown in Figure~\ref{nuerecoil}b.
The SM contribution is of the order of 
1~$\rm{kg ^{-1} day^{-1}}$ at
the typical parameters for reactor experiments and 
detection threshold of 100~keV

\begin{figure}
{\bf (a)}
\centerline{
\epsfig{file=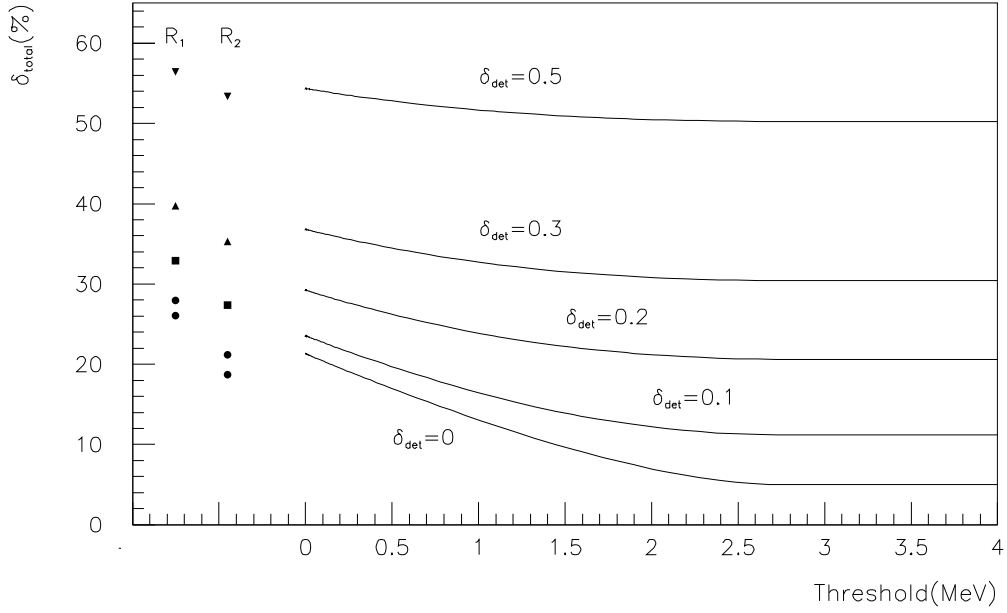,width=14cm}
}
{\bf (b)}
\centerline{
\epsfig{file=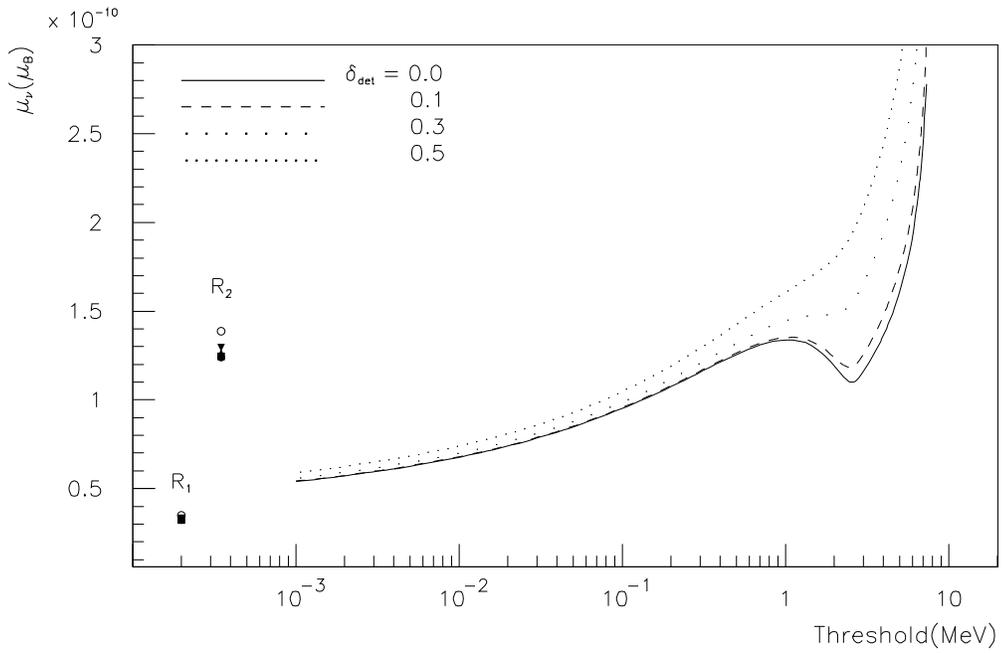,width=14cm}
}
\caption{
Attainable sensitivities in $\nuebar$-e scattering
experiments as a function
of detection threshold for
(a) SM cross section measurements and
(b) MM limits at 90\% CL,
in the case where $\xi$=3~MeV and
$\Delta$=30\%, for different
values of the experimental uncertaintites $\dmeas$.
The different symbols for the ranges
$\rm{R_1}$ and $\rm{R_2}$
correspond to the different $\dmeas$ values.
}
\label{sensit}
\end{figure}

With the conservative but realistic values of 
$\Delta$=30\% and $\xi$=3~MeV,
the attainable total ``1-$\sigma$'' accuracies 
for the SM cross-sections are evaluated 
and depicted in Figure~\ref{sensit}a
as a function of detection threshold 
for different values of 
the measurement error $\dmeas$.
Also shown are sensitivities for the detection
ranges of 
$\rm{R_1}$=5-100~keV and $\rm{R_2}$=0.5-2~MeV,
which correspond to the ranges
of the on-going experiments Kuo-Sheng~\cite{texono} 
and MUNU~\cite{munu}, respectively.
It can be seen that, for the same $\dmeas$,
measurements with a low 
threshold are limited by the uncertainties
in $\rnusp$. An experiment 
optimized for SM cross-section measurements
should focus on the events with higher recoil energy (above 1.5~MeV)
while trying to compensate for the
loss of statistical accuracy with large target mass,
that is, keeping the threshold high without
compromising $\dmeas$.

In an analysis for the magnetic moment effects,
the contributions from $\dmm$
should be taken into consideration.
Since the uncertainties in the evaluation
of $\rnusp$ will
translate into correlated errors of the same sign 
for both $\dsm$ and $\dmm$,
the combined experimental uncertainties 
can be written as
\begin{equation}
\rm{
\dtotal ^2 ~ = ~
\dmeas ^2 ~ + ~ ( \dsm + \dmm ) ^2 ~ .
}
\end{equation}
In other words, an under-estimation of $\rnusp$
would lead to an excess of events after Reactor ON/OFF
subtraction, which can be taken as signatures of 
positive magnetic moments.
The positive signals would be interpreted as  
even bigger values of $\munu$ when 
the {\it same} under-estimated $\rnusp$ is used
to evaluate the magnetic moment.

The attainable MM limits at 90\% confidence level (CL)
can be derived from $\dtotal$, and 
are displayed in Figure~\ref{sensit}b as a function
of threshold and for different values of $\dmeas$.
The sensitivities trend is distinctively different
from that of SM cross-section measurements in Figure~\ref{sensit}a.
The SM cross-section becomes ``background'' to MM searches,
such that in the energy range 
where SM interactions dominate,
the SM uncertainties will get amplified by the 
big SM:MM ratio
in the derivation of magnetic moments.
The MM effects, therefore, should be investigated at
regions where MM is much larger than SM, that is,
at low recoil energies.
The structures at 1-3~MeV for small $\dmeas$ are
due to the sharp transition at $\xi$=3~MeV
in modeling the $\rnusp$ uncertainties.
A more realistic description is that
$\Delta$ would increase continuously as
$\enu$ drops below 3~MeV.

Figure~\ref{sensit}b indicates 
several strategic features in
the experimental search for
neutrino magnetic moments with reactor neutrinos.
In the scenario where $\dmeas$=30\%,
(1) experiments with threshold of $>$1~MeV recoil
energy cannot probe below $10^{-10}~\mub$, 
(2) experiments with range $\rm{R_2}$
cannot probe below $1.2 \times 10^{-10}~\mub$, and 
(3) a sensitive search should be conducted with
as low a threshold as possible and preferably
with a high energy cut-off.  The sensitive region
can approach $3 \times 10^{-11}~\mub$ for the 
$\rm{R_1}$ range. 

\begin{figure}
{\bf (a)}
\centerline{
\epsfig{file=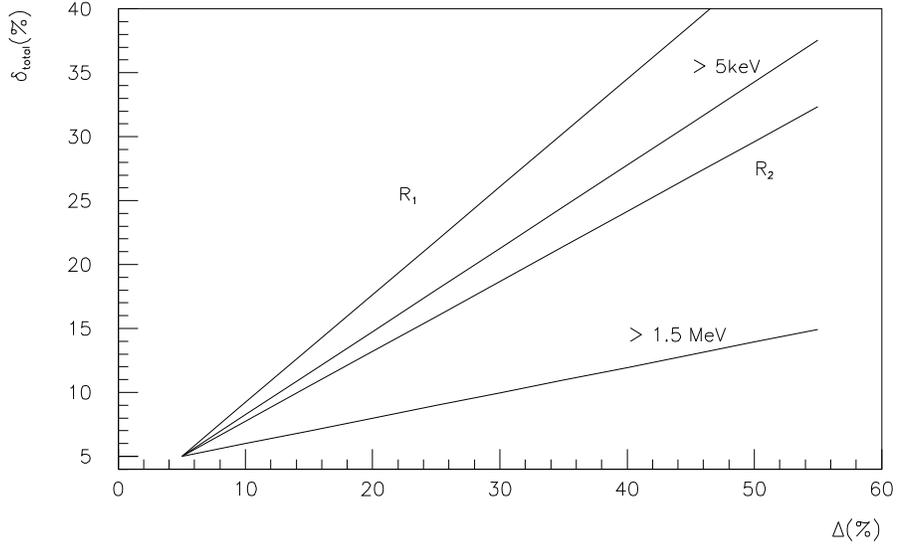,width=14cm}
}
{\bf (b)}
\centerline{
\epsfig{file=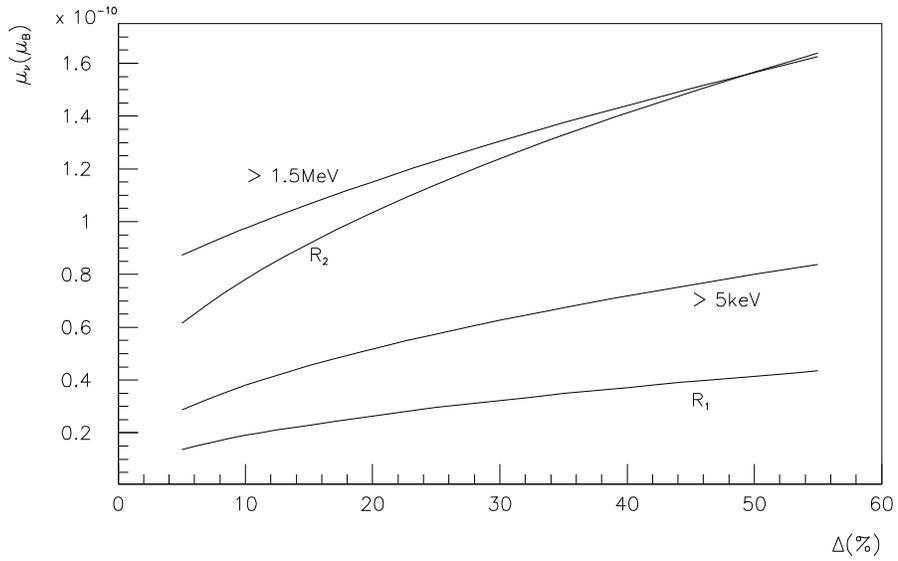,width=14cm}
}
\caption{
Attainable sensitivities in $\nuebar$-e scattering
experiments for
(a) SM cross section measurements and
(b) MM limits at 90\% CL,
in the case where $\xi$=3~MeV and $\dmeas$=0\%,
as a function of $\Delta$, the uncertainty parameter
for the low energy reactor neutrino spectra.
Different contours correspond to 
different detection ranges.
}
\label{delta}
\end{figure}

To investigate the effects of $\Delta$ keeping $\xi$=3~MeV, 
the sensitivities for both SM cross section measurements
and MM limits are shown in Figures~\ref{delta}a and \ref{delta}b,
respectively. The special case of $\dmeas$=0\%
is chosen to characterize a {\it perfect} experiment 
where the sensitivities are limited
only by the uncertainties of $\rnusp$.
The differences in the attainable sensitivities
among the different energy ranges are very distinct 
between the two measurements.
A high threshold value provides best sensitivities
in SM cross-sections while a restricted low energy range
$\rm{R_1}$ is optimal for MM searches.
As depicted in Figure~\ref{delta}a,
a $\dtotal < 10$\% SM measurement is in principle possible
with  a high threshold ($>$1.5~MeV) experiment
even for $\Delta$=30\%. 
In this case, the sensitivities are limited
by experimental uncertainties $\dmeas$ instead.
On the other hand, measurements with low energy data
will require improving $\Delta$ to better than 10\%
to achieve the same sensitivities.

Figure~\ref{delta}b shows that the $\rm{R_1}$-class
of experiments are the least sensitive to
the uncertainties in $\rnusp$, even for very big $\Delta$.
If the entire range of $\rnusp$ can be known to 5\%, such experiments
can probe the region down to $10^{-11}~\mub$.
In contrast, the goals of the $\rm{R_2}$-class
experiments for achieving better than $\rm{10^{-10}~\mub}$
should be complemented by a demonstration of the 
control of the low energy part of the reactor neutrino spectrum 
to the $<$20\% level. 

The effects of the uncertainties in $\rnusp$
on the derived magnetic moment limits were not discussed
in previous work~\cite{reines,kurt,rovno}.
The $\nuebar$-e reactor experiment
in Ref~\cite{reines} had  a threshold of 1.5~MeV
and an uncertainty in the SM cross-section
measurement of $\dmeas$=29\%.
A reanalysis of this experiment
in Ref~\cite{vogelengel}
with improved input parameters
on $\rnusp$ and $\rm{sin ^2 \theta_W}$
gave a positive signature
consistent with the interpretation of a
finite magnetic
moment at $(2-4) \times 10^{-10}~\mub$.
A possibility to mimic the effect
of a neutrino magnetic moment at $2 \times 10^{-10}~\mub$
could be an under-estimation of $\rnusp$ by
$\Delta$=57\% below 3~MeV.
Taking this value of $\Delta$ to be the
characteristic uncertainties  of $\rnusp$
at low energies,
one can infer from Figure~\ref{delta}b
that the attainable sensitivities for $\munu$
for R$_1$- and R$_2$-classes of experiments
are $4 \times 10^{-11}~\mub$ and
$1.6 \times 10^{-10}~\mub$, respectively.
Similarly, measurements from Ref.~\cite{rovno}
were performed in the range of  500~keV to 2~MeV,
and had an experimental uncertainty of $\dmeas$=50\%.
The quoted upper limit of $1.5 \times 10^{-10}~\mub$
at 68\% CL corresponds to a maximum allowed
$\Delta$=65\%.
With future
experiments probing the level of $\rm{\munu < 10^{-10} ~ \mub}$,
it will be necessary to take into account
the uncertainties in $\rnusp$.

It should be emphasized that the  ``attainable sensitivities''
presented in this section are derived from measurements of 
integrated cross-sections within a specified
energy range of electron recoil energy (that is, counting experiments).
In the cases where statistics are abundant enough for
differential cross-section measurements to become possible, 
sensitivities
can be further enhanced by considering the spectral shape.
Nevertheless, the generic conclusions are still valid: 
(a) experiments for SM cross-section measurements should focus on
large ($>$1.5~MeV) recoil energies, where the events are due
mostly to the $\enu > 3~MeV$ which is well-modeled, and
(b) experiments for MM searches should focus on the $\rm{R_1}$-class
energy ranges, where the uncertainties from SM contributions
are minimized, and the $\rm{ 1 \over  T}$ spectral shape
would provide further constraints.

Technically, the $\rm{R_1}$-class
experiments would be similar to those 
for the searches of Cold Dark Matter~\cite{nu2000}.
The new challenges are to control the 
ambient background
in a surface site $-$ {\it and} in the vicinity
of a power reactor core. Detectors with 
high-purity germanium crystals~\cite{texono,russ}
and crystal scintillators~\cite{texono} have
been discussed. An experimental program is being pursued
at the Kuo-Sheng Reactor in Taiwan~\cite{texono}.

\subsection{Neutrino Source: $\nue$-e Scatterings}
\label{sect::source}

Experiments on $\nue$-e scatterings have
been performed at medium energy accelerators~\cite{lanl}.
Sources of $\nue$ from $^{51}$Cr have been produced
for calibrating
the gallium solar neutrino experiments~\cite{snucr}.

\begin{figure}
\centerline{
\epsfig{file=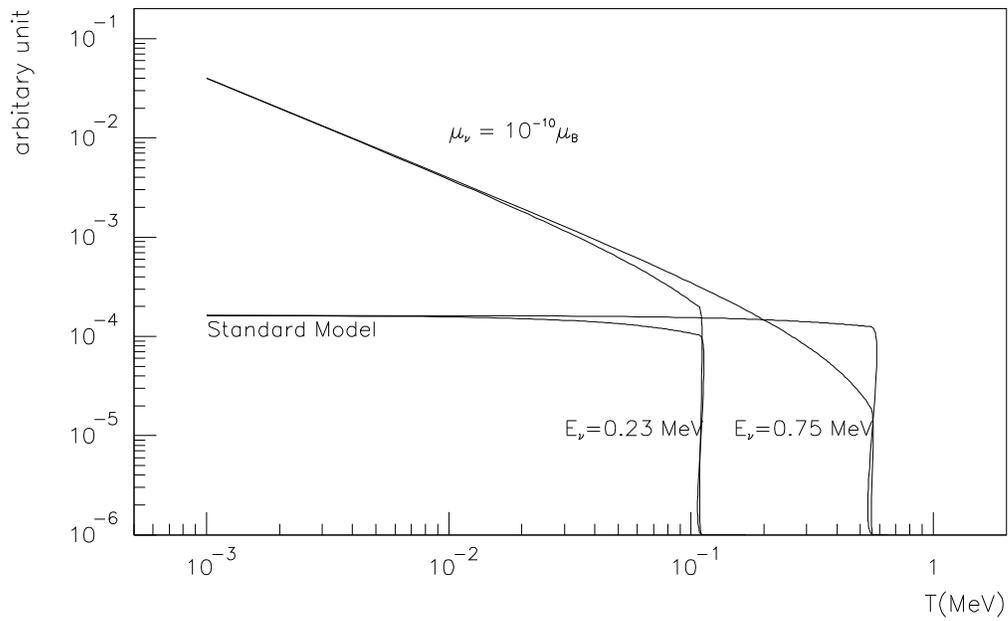,width=16cm}
}
\caption{
Differential cross sections for electron
recoils in $\nue$-e scatterings due to 
SM and MM processes with
$^{51}$Cr and $^{55}$Fe $\nue$ sources.
}
\label{nuediff}
\end{figure}

\begin{figure}
\centerline{
\epsfig{file=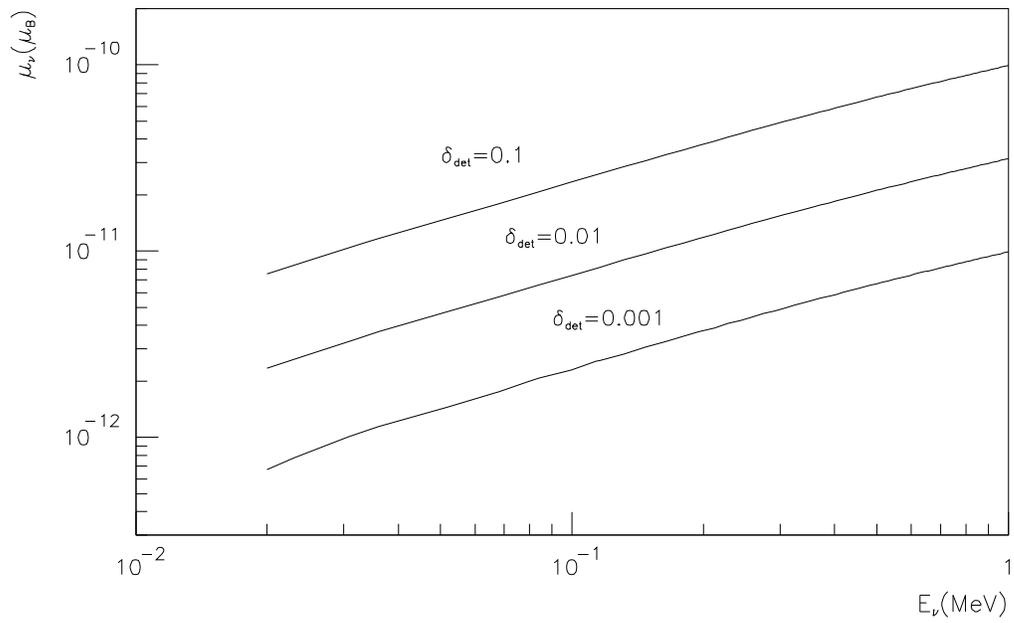,width=16cm}
}
\caption{
The attainable magnetic moment
sensitivities  in $\nue$-e scattering 
experiments
as a function of $\nue$ source energy
at different experimental uncertainties
$\dmeas$.
}
\label{mmsource}
\end{figure}

Studies of neutrino magnetic moments with
artificial neutrino sources have 
been discussed~\cite{nusource,h3source}.
In a similar spirit, the sensitivities
on SM cross section measurements and MM limits using
a $\rm{\nu_e}$ mono-chromatic source are studied.
The differential cross-sections for both
SM and MM at $10^{-10} ~ \mub$ are shown
in Figure~\ref{nuediff} for
two illustrative cases: $^{51}$Cr at 750~keV and
$^{55}$Fe at 230~keV.
The attainable MM limits as a function of
$\enu$ for different $\dmeas$
are shown in Figure~\ref{mmsource}.
The detection threshold for the
recoil electrons is taken to be 1~keV.
The values of $\dmeas$ represent  the combined 
uncertainties due to the experiments
and the measurements of source strength.
For instance, if a 1\% measurement can be made,
the MM sensitivities of $< 10^{-11} ~ \mub$ may 
be probed.

In principle,
experiments with neutrino sources allows
better systematic control and more
accurate ``SOURCE-OFF'' background measurements.
Specific spectral shape for the final-state
measurables can be studied.
For instance, the energy of the final-state
electron spectra in $\nue$-N charged-current interactions
would also be delta-functions,
as considered in the calibration measurements
in the proposed LENS project~\cite{lens}.
An interesting extension to the $\nue$-e
scattering studies is the study of the
``Compton'' edges due to scatterings of
the mono-energy $\nue$, an experimentally cleaner
signature. 

To probe MM sensitivities to the $\rm{10^{-12}~\mub}$
level and beyond, new technologies such
as the various cryogenic detectors with
much lower (100~eV or less) 
detection threshold have to be developed
$-$ a formidable experimental challenge.
The ``neutrino-related-background'' 
at very low energies
will be dominated by the coherent scatterings
on nuclei in reactor neutrino experiments.
The spectra below 2~keV shown in Figure~\ref{nuerecoil}a are
due to coherent scatterings of reactor $\nuebar$
on germanium,
assuming a complete detection of the total recoil energy
(typical ionization yield for germanium at this energy range 
is only about 0.2$-$0.3).
To minimize the contributions
of the SM background of {\it both}
$\nu$-e and $\nu$-N coherent scatterings, 
low energy neutrino sources will be appropriate.
Schemes are considered using 
tritium $\nuebar$ source where
$\rm{\emax=18.6~keV}$~\cite{h3source}.
Reactor neutrinos can still be of use only
if the detectors can provide very good 
event identification capabilities, such as 
the pulse shape discrimination (PSD) techniques, to
differentiate electrons from nuclear recoils.

However,
the statistical accuracy should
also be put into consideration
in realistic experiments.
A neutrino ``point'' source of 1~MCi strength
placed at the center of a spherical
detector of radius 1~m is equivalent
to an exposure to a homogeneous flux of 
$\rm{8.8 \times 10^{11}~cm^{-2} s^{-1}}$,
as compared to that for
typical reactor experiments at
$\rm{10^{13}~cm^{-2} s^{-1}}$.
Coupled with the tremendous efforts and expenses
of producing  the neutrino sources as well
as their finite life-times (for instance,
$\rm{\tau_{\frac{1}{2}}=28~days}$
for $^{51}$Cr), reactor neutrinos still offer
advantages in the study of low energy neutrino
physics.

For completeness and comparison, we
mention that nuclear power reactors
also produce $\nue$~\cite{nuereactor}, 
expected to be predominantly
from $^{51}$Cr and $^{55}$Fe 
via neutron activation of the equipment and building materials,
at the estimated level of about $10^{-3}$~$\nue$/$\nuebar$.
The effects from the small
contaminations of $\nue$ on the measurements
of $\nuebar$ are therefore negligible.
Since the unstable parent isotopes have
relatively long half-lives,
experimental studies with $\nue$ 
can in principle be performed by
studying the transient effects 
{\it after} the reactor is switched OFF, where
the signatures would have the characteristic
half-lives, such as that of 28~days in the 
case of $\nue$'s from $^{51}$Cr.
Such experiment has been considered
for studying the possible
anomalous matter effects of
$\nue$ which may be absent in $\nuebar$~\cite{vannucci}. 

\section{Neutrino Interactions on Nuclei}
\label{sect::nunuclei}

Neutrino cross-sections on nuclei is another subject
which can be studied with reactor neutrinos. The
charged- and neutral-current interactions on
deuteron have been experimentally measured~\cite{nud}, 
while neutral-current excitations have been studied 
theoretically~\cite{nuex}. 
The $\nuebar$N charged-current interactions have
also been discussed in connection to the
detection of low energy $\nuebar$ from the Earth~\cite{terrest}.
It is therefore relevant to study the
attainable accuracies of these cross-sections with
reactor neutrinos under the scenarios mentioned above.

The neutral-current excitation processes:
\begin{equation}
\rm{
\nuebar ~ + ~ N ~ \rightarrow ~ \nuebar ~ + ~ N^* 
}
\end{equation}
have the dependence of
\begin{equation}
\rm{
\sigma ( \enu ) ~ \propto ~  ( ~  \enu ~ - ~ E_{ex} ~ ) ~ ^2
}
\end{equation}
where $\rm{E_{ex}}$ is the threshold
excitation energy. 
It has been observed only in the case of $^{12}$C
with accelerator neutrinos~\cite{karmennuex}.
Theoretical work~\cite{nuexaxial} suggests that 
these cross sections
are sensitive to the
axial isoscalar component of the weak
neutral-current interactions
and the strange quark content of the nucleon.

\begin{figure}
\centerline{
\epsfig{file=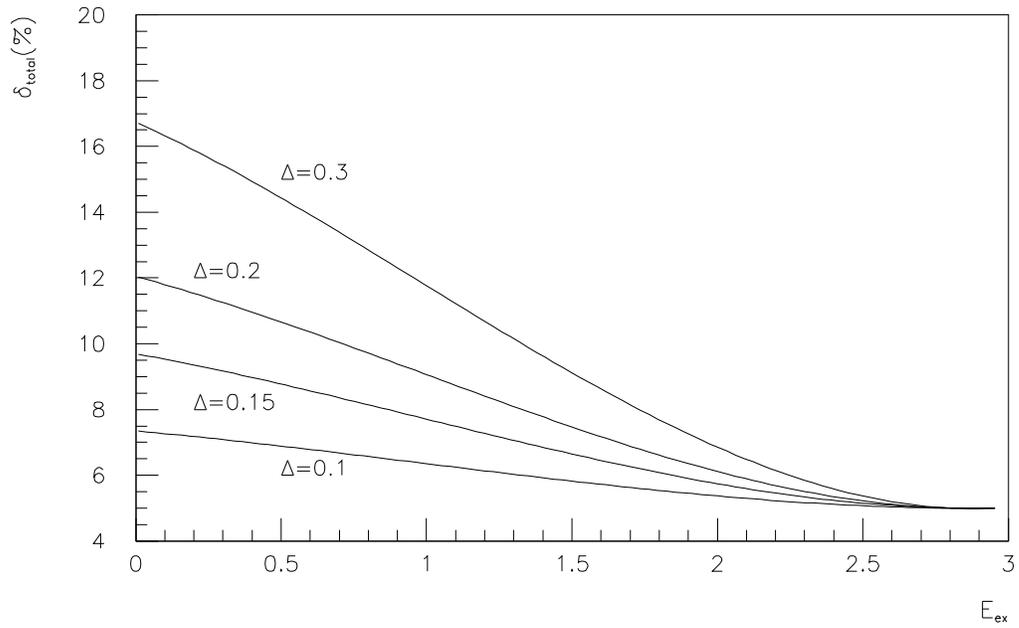,width=16cm}
}
\caption{
Attainable accuracies for reactor-based
neutrino neutral-current excitation experiments
as a function of the nuclear excitation threshold
for different values of $\Delta$
in the cases of 
{\it perfect} experiments ($\dmeas$=0\%).
}
\label{nuexfig}
\end{figure}

The attainable accuracies as a function of
$\rm{E_{ex}}$ at $\dmeas=0$ for different values of $\Delta$
are displayed in Figure~\ref{nuexfig}.
To achieve a 
10\% accuracy in the cross-section measurement 
in the most promising case for the M1 transition
in $^7$Li ($\rm{E_{ex}=448~keV}$), 
it is necessary to 
evaluate the low energy part of $\rnusp$ to better than
16\%.

Neutrino disintegrations on deuteron involve
three-body final states:
\begin{equation}
\rm{
\nuebar ~ + ~ ^2 H  ~ \rightarrow ~
n ~ + ~ n ~ + ~ e^+ ~~  ( E_T = 4.03~MeV ) ~ ,
}
\end{equation}
and 
\begin{equation}
\rm{
\nuebar ~ + ~ ^2 H  ~ \rightarrow ~
\nuebar ~ + ~ p ~ + ~ n ~~  ( E_T = 2.226~MeV ) ~ 
}
\end{equation}
for the charged- ($\nuebar$dCC)
and neutral-current ($\nuebar$dNC)
channels, respectively.
The dependence on
the threshold energy $\rm{E_T}$ is modified to
\begin{equation}
\rm{
\sigma ( \enu ) ~ \propto ~  
\int  \sqrt{E_r} ~ ( \enu - E_{T} - E_r + m ) ~ 
[ ( \enu - E_T - E_r + m )^2 - m^2 ] ^{\frac{1}{2}}  ~ 
d E_r  ~ ,
}
\end{equation}
where $\rm{E_r}$ is the reduced kinetic energy
of the final proton and neutron, and
m=$\rm{m_e}$ and 0 for $\nuebar$dCC
and $\nuebar$dNC, respectively.
There is a sharp increase in 
the cross-section for $\nuebar$dNC
near threshold, such that  
only 0.43\% of the events in a reactor experiment
would originate from $\nuebar$ of $\enu < 3$~MeV.
Accordingly, the attainable accuracies in both channels
are limited only
by the uncertainties of the high energy part of
$\rnusp$, which is about 5\%. 
These uncertainties are better than
the experimental errors~\cite{nud} achieved
at present.

\begin{figure}
\centerline{
\epsfig{file=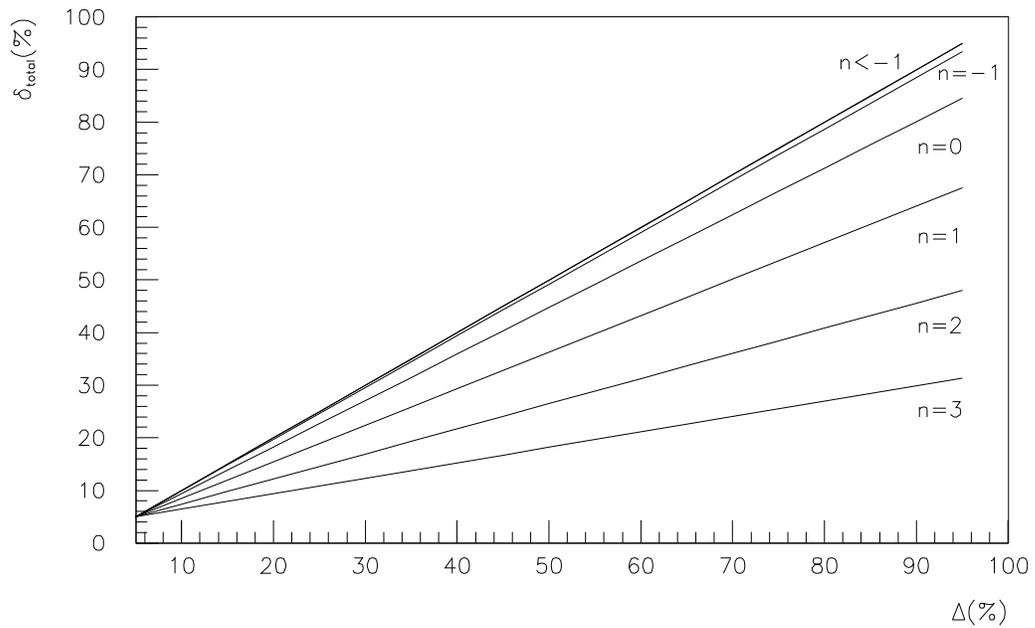,width=16cm}
}
\caption{
Attainable accuracies for cross-section
measurements of reactor
neutrino with nuclei where the energy
dependence is parametrized by Eq.~\ref{eq::nuint},
as a function of index n and 
for various values of
$\Delta$ 
in the cases of 
{\it perfect} experiments ($\dmeas$=0\%).
}
\label{nuint}
\end{figure}

One can extend the studies to the generic
case where the neutrino 
interactions do not have thresholds but
possess an energy dependence parametrized by an index n,
such that
\begin{equation}
\label{eq::nuint}
\rm{
\sigma _{\nu N} ( \enu ) ~ \propto ~  \enu  ^n  ~~~~~ .
}
\end{equation}
The attainable accuracies for an integral
cross-section measurement 
for different values of n
as a function of $\Delta$
are displayed in Figure~\ref{nuint}.
As expected, cross-sections with large n favor
large $\enu$ such that the accuracies approach 
that of $\rnusp$ at high energy.
Interactions with n$\leq -1$, on the other hand,
are dominated by small $\enu$ and the uncertainties
are given by $\Delta$.

As indicated in Figure~\ref{nuerecoil}a,
the coherent scatterings
of low energy neutrinos on nuclei
limit the $\nuebar$-e threshold and therefore
the MM sensitivities in
reactor neutrino experiments.
The corresponding cross sections due to 
the SM and MM processes are~\cite{coher}:
\begin{equation}
\label{eq::cohsm}
\rm{
( \frac{ d \sigma }{ dT } ) ^{coh} _{SM}  ~ = ~
\frac{ G_F^2 }{ 4 \pi }
m_N  [ Z ( 1 - 4 \s2tw ) - N ]^2  
[ 1 -  \frac{m_N T_N }{2 E_{\nu}^2 } ] ~~~~ and
}
\end{equation}
\begin{equation}
\label{eq::cohmm}
\rm{
( \frac{ d \sigma }{ dT } ) ^{coh} _{MM}  ~ \sim ~
\frac{ \pi \alpha _{em} ^2 \munu ^2 }{ m_e^2 }
Z^2
 [ \frac{ 1 - T/E_{\nu} }{T} ]  ~~ , ~~ respectively,
}
\end{equation}
where $\rm{m_N}$, N and Z are the mass, neutron number 
and atomic number of the nuclei, respectively,
and $\rm{T_N}$ is their recoil energy.
Both of the $\sim$N$^2$ and Z$^2$ dependences signify coherence.
This SM interaction is of significance 
in astrophysical processes 
but has not yet been observed in
an experiment due to the extremely small energy
depositions in nuclear recoils.
It dominates over the $\nuebar$-e scatterings
at recoil energy less than $\sim$1~keV.
The integral SM cross-section 
is characterized by n=2 such
that at $\Delta$=30\%, a cross-section measurement
with an accuracy of 15\% can be achieved
in a reactor-based experiment.
The effects of magnetic moments at $10^{-10}~\mub$ 
in the coherent scattering channel are 
relevant only at recoil energy less than 10~eV.

\section{Summary and Discussion}

The strong and positive evidence of neutrino oscillations
implies the existence of neutrino masses and mixings,
the physical origin, structures and 
experimental consequences of which
are still not thoroughly known or understood.

Experimental studies on the neutrino properties
and interactions which may reveal some of these
fundamental questions are therefore of interest and
relevance. Nuclear power reactors remain the most
available and intense sources of neutrinos,
and can contribute to these studies.
The low energy (MeV scale) and
that being related to the first family 
(and therefore allowing the 
possibility of anomalous matter effects)
may favor exotic phenomena to
manifest themselves.
The low energy part of the reactor neutrino
spectra has not been well modeled.
To study detection channels other than 
$\nuebar$ on proton, the $\rnusp$ at low energy
will have
to be worked out and the accuracies shown to be
in control. Future work along this direction
will be of interest.

In this article, we discussed the origins of the 
uncertainties in the modeling of the
low energy reactor neutrino spectra. 
Neutrino emissions from long-lived isotopes as well
as from final states due to 
neutron excitations have to be taken
into account.
We studied how the uncertainties may limit
the sensitivities in measurements of 
reactor neutrino with electrons and nuclei. 
The discrepancies between the results of Ref.~\cite{reines}
and the analysis of Ref.~\cite{vogelengel} can be
explained by the under-estimation of the low energy
part of $\rnusp$ by 57\%.

To optimize the cross-section measurements
for $\nuebar$-e scatterings,
one should focus on the
high energy ($>$1.5~MeV) electron recoil events.
For magnetic moment searches, it would
be best to restrict to the $<$100~keV range
where the effects due to 
the uncertainties of the
Standard Model ``background'' are
mostly decoupled. 
On the other hand, 
experiments which rely on the intermediate
energy range  of 0.5 to 2 MeV
will be limited
in sensitivities in both cross-section measurements
and magnetic moment searches $-$ unless the 
precision of the low energy reactor neutrino
spectra is demonstrated.
An experimental program adopting these strategies
is being pursued
at the Kuo-Sheng Power Reactor Plant~\cite{texono}.
A high-purity germanium detector is employed to optimize
the detector threshold while CsI(Tl) crystal scintillators
are adopted to study the high energy events 
taking advantage of their many potential merits~\cite{prospects}
such as large available mass and yet being compact in 
size.

Artificial neutrino sources are 
attractive alternatives which may
offer better systematic control.
The event
rates tend to be less than those with
reactor neutrinos, unless both the source
and the detector can be made very compact.
For similar uncertainty levels on
the source strength,
the attainable sensitivities
in  both cases are comparable.
To achieve the $\rm{10^{-12}~\mub}$ range
and beyond for magnetic moment searches,
very low energy neutrino sources such that tritium 
is more appropriate, complemented with 
new detector technology
with the range of 10-100~eV threshold.


\section{Acknowledgements}

The authors would like to thank P. Vogel for
stimulating discussions and information on
the evaluations of reactor neutrino spectra.
We are grateful to S.~Pakvasa, F.~Vannucci,
Z.Y.~Zhou and J.~Li for their valuable input.
This work was supported by contracts
NSC~89-2112-M-001-056 and NSC~90-2112-M-001-037
from the National Science Council, Taiwan.
The support of H.B.~Li is from  contracts
NSC~90-2112-M002-028 and MOE~89-N-FA01-1-0
under P.W.Y.~Hwang.\\

\clearpage

\clearpage

\clearpage

\clearpage

\clearpage

\end{document}